\documentclass[12pt]{article}
\usepackage[colorlinks=true, citecolor=blue]{hyperref}
\usepackage{enumerate}
\usepackage{mathtools, amssymb, latexsym, amsmath, amsfonts, amsthm, xcolor,braket}
\usepackage{array, fancyhdr, bm, listings}
\usepackage{algorithm,comment}
\usepackage{algorithmic}
\usepackage{subcaption}
\usepackage{thmtools, thm-restate,physics}
\declaretheorem{theorem}
\usepackage{ bbold }

\theoremstyle{plain}
\newtheorem{lemma}[theorem]{Lemma}

\newtheorem{corollary}[theorem]{Corollary}

\theoremstyle{definition}

\newtheorem{remark}[theorem]{Remark}



\newcommand\sym{\mathcal{S}_n}

\providecommand{\keywords}[1]{\textit{Keywords:} #1}
\title{Discrete Quantum Walks on the Symmetric Group}

\date{}

\author{Avah Banerjee}
\newcommand{\FormatAuthor}[3]{
\begin{tabular}{c}
#1 \\ {\small\texttt{#2}} \\ {\small #3}
\end{tabular}
}
\author{
\begin{tabular}[h!]{lcr}
   \FormatAuthor{Avah Banerjee}{banerjeeav@mst.edu}{Missouri S\&T}
\end{tabular}
}

\begin{document}

\maketitle

\begin{abstract}
The theory of random walks on finite graphs is well developed with numerous applications.
In quantum walks, the propagation is governed by quantum mechanical rules; generalizing random walks to the quantum  setting.
They have been successfully applied in the development of quantum algorithms. In particular, to solve problems that can be mapped to searching or property testing on some specific graph. 
In this paper we investigate the discrete time coined quantum walk (DTCQW) model using tools from non-commutative Fourier analysis.
Specifically, we are interested in characterizing the DTCQW on Cayley graphs generated by the symmetric group ($\sym$) with appropriate generating sets. The lack of commutativity makes it challenging to find an analytical description of the limiting behavior with respect to the spectrum of the walk-operator.
We determine certain characteristics of these walks using a path integral approach over the characters of $\sym$.

\vspace{0.5cm}
\keywords{Quantum Walks, \and Symmetric Group, \and Non-commutative Fourier analysis}
\end{abstract}

\section{Introduction}
The phenomenon of random walks on graphs has been widely studied and applied to a wide verity of problems in computational sciences. In particular they have been instrumental in developing randomized and approximation algorithms \cite{lovasz1993random}. More recently higher dimensional analogue of random walks (over simplicial complexes) have been proposed \cite{lubotzky2014ramanujan}.
Propagation properties of random walks can be characterized by a Markov chains.
Hence the walk is amenable to characterization using  methods from spectral graph theory\cite{godsil2001algebraic}. 

Unlike \emph{classical} random walks a \emph{quantum walk} propagates using the principle of quantum mechanics. Few difference of note include - 1) Instead of real probabilities the state of the walk is specified by complex probability amplitudes\footnote{However, in some case if the amplitudes are constrained to be in $\mathbb{R}$, working with them becomes slightly simpler.}. 2) The random (walk) coin is now replaced by a unitary transformation. The unitary evolution ensures the walk is reversible\footnote{For open systems the walk operator need not be unitary. Interspersing walking with measurements also leads to non-unitary dynamics\cite{kendon2007decoherence}}.
2) Propagation of the walk generates a superposition state overs all possible positions available to the walker. 
3) Finally, we can sample the positions by applying suitable measurements on the state of the walker.

There are various (somewhat equivalent) models of quantum walks.
Study of quantum walks has a long history, going back to the early works of Feynman, Meyer, Aharonov, Gutmann and others \cite{aharonov1993quantum, farhi1998quantum, meyer1996quantum}.
The hope is that quantum walk can emulate the success of random walk in the development of classical algorithms in developing quantum algorithms.  Quantum or classical walk\footnote{Henceforth we will refer to classical random walk simply as classical walk.} has been primarily used as a generative models for probability distributions. 
Hence, two of the most important properties to study are the kind of distributions they can generate and their converging behavior. In general, quantum walks do not converge to a stationary distribution. However, their time-averaged distribution (introduced later) does converge.
Quantum walk has been shown to generalize Grover's diffusion based search on graphs. It has been used to obtain currently best known quantum algorithms for certain problems. Most notable among them are element distinctness, triangle finding, faster simulation of Markov chains, expansion testing etc. \cite{magniez2011search, ambainis2008quantum, apers2020expansion}. 

\paragraph{Results.}
In this paper we focus on a discrete time model of quantum walk. The model we study originated in the seminal paper by Aharonov et. al.\cite{aharonov2001quantum}. The model is also referred to as discrete time coined quantum walk (DTCQW). We study walks on Cayley graphs of the symmetric group with appropriate generating sets. As our main result, we derive a path-integral type expression for the amplitudes using non-commutative Fourier analysis.
We show that if the generating set is closed under conjugation then the distribution is uniform over the conjugacy classes if the initial state of the walker in the coin-basis is the uniform superposition state. Unfortunately this result does not hold when starting from an arbitrary basis state.
Additionally, we study the characteristics of the Hadamard walk on the graph generated by $\{(12),(1\cdots n)\}$. This graph was chosen due its simplicity as well as being less ``expander like''. 

\section{Preliminaries}
\subsection{Cayley Graphs}
\begin{figure}[h]
	\includegraphics[width=5cm]{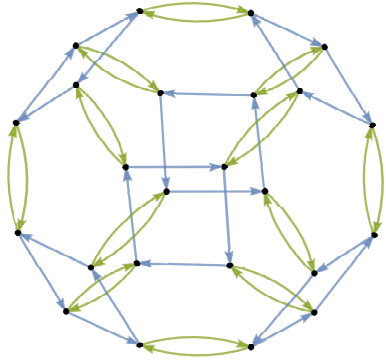}
	\centering
	\caption{The graph $\Gamma_4$. Edges corresponding to the element $(12)$ (resp. $(1\cdots n)$) are colored green (resp. blue). It has $24$ vertices and has a diameter of $6$.}
\label{fig: gamma n}
\end{figure}
Let $(G, \circ)$ be any finite group  and $S$ be a generator of $G$. We take $|G| = N$ and $|S| = d$.
The Cayley graph of the pair $(G, S)$ is a directed graph $\Gamma$ defined as follows. 
The vertex set $V(\Gamma) = G$.
The edge set is defined as $$E(\Gamma) = \{(g,h), g, h \in G\mid\ \mbox{$\exists s \in S$ such that $h = g\circ s$}\}.$$
Henceforth, we omit the ``$\circ$" and simply write $g\circ h$ as $gh$, where $g, h \in G$.
If $S$ is closed under inverse, that is $s \in S \implies s^{-1} \in S$ then $\Gamma$ is undirected. 
We use $\mathbb{e}$ to denote the identity element of $G$.
If $\mathbb{e} \not \in S$ then $\Gamma$ does not have any self-loops.
Clearly $\Gamma$ is $d$-regular. This allows for a reversible walk operator, which is a requirement for unitary quantum evolution.
In this paper we associate $G$ with the symmetric group ${\cal S}_n$ of all $n$-permutations. Some typical generators of $G$ are - the set of all transpositions, $\{(12),(13),\ldots, (1n)\}, \{(ij),(1\cdots n)\}$ where $gcd(|i-j|,n)=2$ etc.
Later, we will study the Cayley graph, denoted as $\Gamma_{n}$, with respect to the last generator (specifically $\{(12),(1\cdots n)\}$). Figure \ref{fig: gamma n} shows $\Gamma_{4}$. For $n \ge 3$, $\Gamma_{n}$ is directed with in and out-degree of two. The element $\mu = (12)$ is of order 2 and hence the pair of edges $(g,g\mu)$ and $(g\mu,g)$ could be taken together as an undirected edge.
These edges form perfect matchings.
On the other hand, the element $\sigma = (1\cdots n)$ creates directed $n$-cycles. Lastly, we say $S$ is \emph{conjugate invariant} if it
is a union of one or more conjugacy classes. For example, $S = $ set of all transpositions.

\subsection{Formal Description of DTCQW}
Physically, a particle with some internal degrees of freedom moves in superposition, as it propagates on $G$. The state of such a particle at any moment is described a vector in the Hilbert space ${\cal H}$ with a basis set $\{\ket{s,g}\mid\ \mbox{$s \in S$ and $g \in G$} \}$ (standard basis). Thus we can write ${\cal H} = {\cal H}_G\otimes{\cal H}_S$.
The space ${\cal H}_G$ describes the position of the particle over the group elements (alternatively over the vertices of  $\Gamma$).
${\cal H}_S$ is the coin (chiral) space, which describes the state of particle's internal degrees of freedom (sometimes referred to as the particles \emph{chirality}).
One step of the walk consists of applying the two unitaries $C_S \otimes I_G$ and $\Lambda$ in succession.
We first apply the coin operator $C \otimes I_G$  which acts trivially on ${\cal H}_G$. This transforms the chiral state of the particle.
Although, there are no particular restrictions on the unitary $C$, in this work, we mainly consider the case when $C$ is the Grover operator.
Next, we apply the shift operator $\Lambda$ which acts on the total space $\cal H$, and performs a conditional shift of the particle's position based on its current chiral state in ${\cal H}_S$.
Together, each step of the walk consists of applying the unitary $U = \Lambda (C \otimes  I)$ to the current state. We describe $C$ and $\Lambda$ next.
\subsubsection{Coin operators}
For $d \ge 3$ the Grover operator $D$ (reflection about the mean) is defined as follows. $D$ is also commonly known as the diffusion operator.  It is defined as: $D = 2\ketbra{\psi}{\psi}-I$, where $\ket{\psi} = \frac{1}{\sqrt{d}}\sum_{s \in S}\ket{s}$ is the uniform superposition over the basis states. $D$ acts only on the coin space ${\cal H}_S$. 
 Let $\delta_{ij}$ be the Kronecker delta function.
In the matrix notation $(i,j)^{th}$ entry of $D$ is given by:
 $D_{ij} = \delta_{ij}a + \left(1-\delta_{ij}\right)b$
 where $a = \frac{2}{d}-1$ and $b = \frac{2}{d}$.
When $|S| = 2$ we consider the Hadamard operator $H = \frac{1}{\sqrt{2}}\begin{bmatrix}1 & 1 \\ 1 & -1\end{bmatrix}$ or the operator $\frac{I+iX}{\sqrt{2}}$. Here $X$ is the controlled not gate. It has been shown that the propagation of the walk on the line when $C = \frac{I+iX}{\sqrt{2}}$ is symmetric \cite{lipton2014quantum} as opposed to $H$ which has a heavy tail on one side. 

\subsubsection{The $\Lambda$ operator}
The shift operator $\Lambda = \sum_{s \in S, g \in G}\ketbra{s,gs}{s,g}$. In literature it is sometimes referred to as the move operator to distinguish it from some of its extensions.
$\Lambda$ sends the walker with internal chiral state $\ket{s}$ and at position $g$ along the edge $s$ to $gs$.
In the matrix form, $\Lambda$ is a $dn \times dn$ block diagonal matrix with $d$ blocks. There is a block corresponding to each $s \in S$. The block corresponding to $s$ is the $n \times n$ permutation matrix associated with the action of $s$ on $G$.
A more general version of $\Lambda$ also permutes the basis in ${\cal H}_S$. Specifically, $\Lambda_{\pi} = \sum_{s \in S, g \in G}\ketbra{\pi(s),gs}{s,g}$. In the case of the grid graph, $\pi$ performing a directional flip ($\ket{\uparrow} $ to $\ket{\downarrow}$ and $\ket{\leftarrow} $ to $ \ket{\rightarrow}$ and vice versa), gives rise to the so-called flip-flop walk \cite{shenvi2003quantum}.

\subsubsection{Initial states and evolution}
We use $\ket{\psi_t} = \alpha_{s,t}(g)\ket{s,g}$ to denote the state of the walker after $t$ steps. $\ket{\psi_0}$ is the initial state. We can write,
$$\ket{\psi_t} = U^t \ket{\psi_0}$$
Then the probability of observing a particle at $g$ when measured on the standard basis is
$$P_t[g\mid \psi_0] = \sum_s{|\alpha_{s,t}(g)|^2}$$
Since $U$ is unitary, $\ket{\psi_t}$ shows periodic property \cite{aharonov2001quantum} as long as $\ket{\psi_0}$ is not an eigenvector of $U$.
In general $P_t$ does not converge.
However the time averaged distribution (defined below) does.
$$\overline{P}_T[g\mid \psi_0] = \frac{1}{T}\sum_{t=0}^{T-1}P_t[g\mid \psi_0]$$
$\overline{P}_T$ can be interpreted as the expected value of the distribution $P_t$ when $t$ is selected uniformly at random from the set $\{0,\ldots,T-1\}$. If the amplitudes are all real then to study the convergence of $\overline{P}_T$ it suffices to study the amplitudes only. Let $\pi[\ |\psi_0]$ be the limiting distribution of the walk starting from the initial state $\ket{\psi_0}$. Convergence is measured via the total variation distance $\parallel P - \pi\parallel =  \sum_{g}|P[g]-\pi[g]|$. Various convergence parameters have been introduced in the literature. Notable among them is the \emph{mixing time} of the walk. The mixing time itself can be defined in several way. We use the definition from \cite{aharonov2001quantum} which can be thought of as the average mixing time.
$$M_{\epsilon} = \min \{t\mid\ \mbox{$\forall T \ge t, \ket{s,g}: \parallel\overline{P}_T[\ \mid \ket{s,g}] - \pi[\ \mid \ket{s,g}]\parallel$}\}$$


\section{Previous Work}
In their seminal paper \cite{aharonov2001quantum} Aharonov et. al. gave several results on DTCQWs. They characterized the convergence behavior of  walks on abelian groups. They show that the time averaged distribution converges to the uniform distribution whenever the eigenvalues of $U$ are all distinct. They also gave an $O(\frac{n\log n}{\epsilon^3})$ upper bound on the mixing time for $\mathbb{Z}_n$ (the cycle graph). Some lower bounds were also proved in terms of the graph's conductance.
Following their introduction, DTCQW has been studied for several graph families.
Nayak and Vishwanath \cite{nayak2000quantum}  gave a detailed analysis for the line using Fourier analysis.
They were able show that the Hadamard walk mixes almost uniformly with only $O(t)$ steps. Giving a quadratic speedup over its classical counterpart.
Moor and Russell \cite{moore2002quantum}  analyzed the Grover walk on the Cayley graph of $\mathbb{Z}^n_2$ (a.k.a the hypercube). They show an instantaneous mixing time of $O(n)$. Again, this beats the classical $\Omega(n \log n)$ bound.
Acevedo and Gobron studied quantum walks for certain Cayley graphs and in particular gave a several results for graphs generated by free groups \cite{acevedo2005quantum}.
D’Ariano et. al. investigated the case where the group is virtually abelian \cite{d2016virtually}. Virtual abelianity allowed them to reduced the problem to an equivalent one on an abelian group with a larger chiral space dimension and use the Fourier method of \cite{nayak2000quantum}. 
More recently, DTCQW has been studied for the Dihedral group $D_n$ by \cite{dai2018discrete} Dai et. al. Since, $D_n$ is isomorphic to the semi-direct product  $\mathbb{Z}_n \rtimes \mathbb{Z}_2$; (again) the Fourier approach introduced in \cite{nayak2000quantum} carries over. Using which authors gave spectral decomposition of $U$ for the Grover walk. A detailed survey about various types of quantum walks including DTCQW can be found in \cite{venegas2012quantum} and the reference therein. A survey of DTCQW on Cayley graphs can be found in \cite{knittelquantum}.

Finally, we  mention the continuous time quantum walk model studied in \cite{gerhardt2003continuous} by Gerhardt and Watrous.  In the continuous setting the walk operator $U = e^{iAt}$ is a Hamiltonian determined by the adjacency operator of the Caley graph. When $S$ is the set of transpositions  they show, the time averaged distribution is far from the uniform distribution. They explicitly calculate the probability of reaching a $n$-cycle starting from $\mathbb{e}$ by expressing the eigenstates of $U$ using the characters of  ${\cal S}_n$. Unfortunately, in the DTCQW model an analogous description of $U$ seems elusive.

\section{Results via Representation Theory}
We use representation theory to express the amplitudes $\alpha_{s,t}(g)$ using a sum over the irreducible characters. 
Let $\ket{\psi_0}$ be the initial state of the walk.
After $t$ steps the state is $\ket{\psi_t}$ where,
$$\ket{\psi_t} = \sum_{s,g}\alpha_{s,t}(g)\ket{s,g}$$
Since $\alpha_{s,t}(g)$'s are functions from $G$ to $\mathbb{C}$ we can apply non-commutative Fourier transformation to get their duals:

\begin{align}\label{eq: fourier}
    \hat{\alpha}_{s,t}(\rho) = \sum_{g \in G}\alpha_{s,t}(g)\rho(g)
\end{align}
for every $\rho \in \hat{G}$, the set of all irreducible representations of $G$. Where $\rho : G \to GL(V)$ is a homomorphism from $G$ to the space of linear maps on the vector space $V$ satisfying the following.
For all $g, h \in G$, $\rho(g)\rho(h) = \rho(gh)$ and $\rho(\mathbb{e}) = I$.
We denote by $d_{\rho}$, the dimension of $V$, as the dimension of $\rho$.
The \emph{character} of a representation $\rho$ is defined as $\chi_{\rho}(g) = tr(\rho(g))$. Here $tr()$ is the trace operator. Following properties of $\chi_{\rho}$ will be useful: 
\begin{enumerate}
    \item $\chi_{\rho}(\mathbb{e}) = d_{\rho}$
    \item $\forall g,h \in G:\ \chi_{\rho}(gh) = \chi_{\rho}(hg)$ (cyclic property)
    \item $\forall g,h \in G:\ \chi_{\rho}(hgh^{-1}) = \chi_{\rho}(g)$ ($\chi_{\rho}$ is constant over the conjugacy classes)
    \item $\chi_{\rho}(g^{-1}) = \chi_{\rho}(g)^\dagger$ ($A^\dagger$ is the adjoint of the operator $A$)
\end{enumerate}
Proof of the above relations directly follows from the definition of $\chi_\rho$. For further information and introduction to representation theory, especially in the context of random walks, we refer the reader to the monograph by Diaconis \cite{diaconis1988group}.
The book by Terras \cite{terras1999fourier} gives a comprehensive introduction to non-commutative Fourier analysis.

\paragraph{A recurrence for $\alpha_{s,t}(g)$.}
Let $\ket{\psi_{t}'} = (D \otimes I)\ket{\psi_t}$ and $\ket{\psi_{t+1}} = \Lambda\ket{\psi_t'}$ so that $\ket{\psi_{t+1}} = U\ket{\psi_t}$.
Applying the Grover operator $D$ on the basis states in ${\cal H}_S$ we get,
$$\ket{s} \to a\ket{s}+\sum_{s'\in S, s \ne s'}b\ket{s'}$$
This gives $\ket{\phi_t'}$, the intermediate state just after applying the coin operator.
\begin{equation}
    \ket{\psi_t'}=\sum_{s,g}\alpha_{s,t}(g)\left(a\ket{s}+\sum_{s'\in S, s \ne s'}b\ket{s'}\right)\ket{g}
\end{equation}
After applying $\Gamma$ we get the nest state after completing a full step of the walk.
\begin{align*}
    \ket{\psi_{t+1}}&=\sum_{s,g}\alpha_{s,t}(g)\left(\sum_{s'\in S,s'\ne s}b\ket{s',gs'}+a\ket{s,gs}\right) =\sum_{s,g}\left(a\alpha_{s,t}(gs^{-1})+b\sum_{s' \in S, s'\ne s}\alpha_{s',t}(gs^{-1})\right)\ket{s,g}
\end{align*}
This gives a recurrence relation for the amplitude after $t$ steps:
\begin{align*}
    \alpha_{s,t}(g)=a\alpha_{s,t-1}(gs^{-1})+ b\sum_{s' \in S, s'\ne s}\alpha_{s',t-1}(gs^{-1})
\end{align*}
Now we expand Eq. \ref{eq: fourier}, giving
\begin{align}\label{eq: ft}
 \nonumber    \hat{\alpha}_{s,t}(\rho) &= \sum_{g \in G}\alpha_{s,t}(g)\rho(g) = \sum_{g \in G}\left(a \alpha_{s,t-1}(gs^{-1})+b\sum_{s \ne s'}\alpha_{s',t-1}(gs^{-1})\right)\rho(g)\\
     & = \sum_{g \in G}\left(a \alpha_{s,t-1}(g)+b\sum_{s \ne s'}\alpha_{s',t-1}(g)\right)\rho(gs) = \left(a\hat{\alpha}_{s,t-1}(\rho)+b\sum_{s\ne s'}\hat{\alpha}_{s',t-1}(\rho)\right)\rho(s)
\end{align}
Due to the dependence on $\rho(s)$ the above recurrence does not have a closed form solution. However, $\alpha_{s,t}(g)$ can be expressed as a sum of characters.
We derive this next.

\begin{lemma}\label{thm: main -1}
Given a Grover operator acting on ${\cal H}_S$ and the initial state $\ket{\psi_0} = \frac{1}{\sqrt{d}}\sum_{s}\ket{s,\mathbb{e}}$ we have for $t >  0, d > 2$:
\begin{align}\label{eq: main thm}
    \alpha_{s,t}(g) = \frac{1}{\sqrt{d}N}\sum_{k=0}^{t-1}(a^{t-k-1}b^k)\left(\sum_{\rho \in \hat{G},r \in R_{k,t, s}}d_{\rho}\chi_{\rho}(g^{-1}r)\right). 
\end{align}
where every $r \in R_{k,t,s}$ has a generating sequence of the following form: 
\begin{align*}
   r=  \begin{cases}
    s^t \hspace{3cm} \mbox{if $k = 0$}\\
    s_{k}^{p_k}s_{k-1}^{p_{k-1}}\ldots s_{1}^{p_1}s \hspace{1cm} \mbox{otherwise}
    \end{cases}
\end{align*}
 satisfying - 1) $\forall i \in \{0,\ldots,k-1\}, s_i \ne s_{i+1} (s_0 = s)$ and 2) $\sum_{i}p_i = t-1$. 
\end{lemma}

\begin{proof}
We prove this by induction on $t$.
For the base case we take $t = 1$.
From Eq. \ref{eq: ft} we get:
\begin{align*}
    \hat{\alpha}_{s,1}(\rho) &= \left(a\hat{\alpha}_{s,0}(\rho)+b\sum_{s\ne s'}\hat{\alpha}_{s',0}(\rho)\right)\rho(s) = \left(a\sum_{g' \in G}\alpha_{s,0}(g')\rho(g')+b\sum_{s\ne s'}\sum_{g' \in G}\alpha_{s',0}(g')\rho(g')\right)\rho(s)\\
    &= \left(a\alpha_{s,0}(\mathbb{e})\rho(\mathbb{e}) + b\sum_{s\ne s'}\alpha_{s',0}(\mathbb{e})\rho(\mathbb{e})\right)\rho(s) = \frac{\rho(s)}{\sqrt{d}}(a+ (d-1)b) = \frac{\rho(s)}{\sqrt{d}}
\end{align*}
The inverse Fourier transform of $\hat{\alpha}_{s,t}$ is given by \cite{diaconis1993comparison}:
\begin{align*}
    \alpha_{s,t}(g) &= \frac{1}{N}\sum_{\rho \in \hat{G}}d_{\rho}Tr(\rho^{\dagger}(g)\hat{\alpha}_{s,t}(\rho))
\end{align*}
For $t= 1$ we get:
\begin{align*}
    \alpha_{s,1}(g) &= \frac{1}{N\sqrt{d}}\sum_{\rho \in \hat{G}}d_{\rho}Tr(\rho(g^{-1}s))
\end{align*}
For the inductive case, assume Equation. \ref{eq: ft} holds upto $t-1$. Let $\frac{1}{N\sqrt{d}} = \beta$. Then,
\begin{align*}
    \alpha_{s,t}(g) &= a\alpha_{s,t-1}(gs^{-1})+ b\sum_{s'\ne s}\alpha_{s',t-1}(gs^{-1})\\
    &= \beta\sum_{k=0}^{t-2}(a^{t-k-1}b^k)\sum_{\rho, r \in R_{k,t-1,s}}d_{\rho}\chi_{\rho}(sg^{-1}r) +\beta\sum_{s'\ne s}\sum_{k=0}^{t-2}(a^{t-k-2}b^{k+1})\sum_{\rho, r \in R_{k,t-1,s'}}d_{\rho}\chi_{\rho}(sg^{-1}r)\\
    &= \beta\sum_{k=0}^{t-2}(a^{t-k-1}b^k)\sum_{\substack{\rho,\ r \in R_{k,t,s}\\ r = ps,\ p \in R_{k,t-1,s}}}d_{\rho}\chi_{\rho}(g^{-1}r)\\
    &+  \beta\sum_{k=0}^{t-2}(a^{t-k-2}b^{k+1})\sum_{s'\ne s}\sum_{\substack{\rho,\ r \in R_{k+1 ,t,s}\\ r=qs,\ q \in R_{k,t-1,s'}}}d_{\rho}\chi_{\rho}(g^{-1}r)
\end{align*}
Where the second equality follows from the cyclic property of characters and rearranging the sums in the second term. Substituting $k+1$ for $k$ in the above and rearranging the summations in the second term we get 
\begin{align}\label{eq: two part}
  \nonumber  \alpha_{s,t}(g) &= \beta\sum_{k=0}^{t-2}(a^{t-k-1}b^k)\sum_{\substack{\rho \in \hat{G} \\ r \in P}}d_{\rho}\chi_{\rho}(g^{-1}r)\\
    &+  \beta\sum_{k=1}^{t-1}(a^{t-k-1}b^{k})\sum_{s'\ne s}\sum_{\substack{\rho \in \hat{G}\\ r \in Q_{s'}}}d_{\rho}\chi_{\rho}(g^{-1}r)
\end{align}
\ifx false
\begin{align*}
  \beta\sum_{s'\ne s}\sum_{k=0}^{t-2}(a^{t-k-2}b^{k+1})\sum_{\substack{\rho,\ r \in R_{k,t-1}\\ s'\ne s'', r=ps'',r'=rs'}}d_{\rho}\chi_{\rho}(g^{-1}r's)
\end{align*}

We note that,
\begin{align}\label{eq: path sum part 1}
    \sum_{\substack{\rho,\ r' \in R_{k,t}\\ r' = rs}}d_{\rho}\chi_{\rho}(g^{-1}r's) + \sum_{s\ne s'}\sum_{\substack{\rho,\ r' \in R_{k,t}\\ r=ps', r'=rs'}}d_{\rho}\chi_{\rho}(g^{-1}r's) = \sum_{\substack{\rho,\ r \in R_{k,t}\\ r = ps \vee\ (r = qs's' \wedge\ s'\ne s)}}d_{\rho}\chi_{\rho}(g^{-1}rs)
\end{align}
now,
\begin{align}\label{eq: path sum part 2}
    \sum_{s'\ne s}\sum_{k=0}^{t-2}(a^{t-k-2}b^{k+1})\sum_{\substack{\rho,\ r \in R_{k,t-1}\\ s'\ne s'', r=ps'',r'=rs'}}d_{\rho}\chi_{\rho}(g^{-1}rs's) = \sum_{k=2}^{t-1}(a^{t-k-1}b^{k})\sum_{\substack{\rho,\ r \in R_{k,t}\\ s'\ne s'', r=ps''s'}}d_{\rho}\chi_{\rho}(g^{-1}rs)
\end{align}
\fi
Where,
\begin{align*}
    P &= \{r \in R_{k,t,s}\mid\ \exists p \in R_{k,t-1,s}\ r = ps\}\ \mbox{and} \\
    Q_{s'} &= \{r \in R_{k,t,s}\mid\ \exists q \in R_{k-1,t-1,s'}\ r = qs \wedge\ s'\ne s  \}
\end{align*}
Since $R_{k,t,s} = P \cup \left(\bigcup_{s \ne s'} Q_{s'}\right)$ we can combine the two terms in Eq. \ref{eq: two part}  to get,
\begin{align*}
    \alpha_{s,t}(g) = \beta\sum_{k=0}^{t-1}(a^{t-k-1}b^k)\sum_{\substack{\rho \in \hat{G},\ r \in R_{k,t,s}}}d_{\rho}\chi_{\rho}(g^{-1}r)
\end{align*}

\end{proof}

\begin{theorem}\label{thm: main}
Defining,
\begin{align} \label{eq: pathcounts}
    \#_{k,t,s}(g) = |\{r \in R_{k,t,s}\mid r = g\}|
\end{align}
we have
$$\alpha_{s,t}(g) = \frac{1}{\sqrt{d}}\sum_{k=0}^{t-1}(a^{t-k-1}b^k)\#_{k,t,s}(g)$$
\end{theorem}

\begin{proof}
Recall,
\begin{align*}
    \sum_{\rho \in \hat{G}}d_{\rho}\chi_{\rho}(g) = \begin{cases}
    N \hspace{1cm} \mbox{$g = \mathbb{e}$}\\
    0 \hspace{1cm} \mbox{otherwise}
    \end{cases}
\end{align*}
Substituting this in Eq. \ref{eq: main thm} we have,
\begin{align*}
     \alpha_{s,t}(g) &= \frac{1}{\sqrt{d}N}\sum_{k=0}^{t-1}(a^{t-k-1}b^k)\sum_{\rho \in \hat{G},\ r \in R_{k,t,s}}d_{\rho}\chi_{\rho}(g^{-1}r) \\ &= \frac{1}{\sqrt{d}N}\sum_{k=0}^{t-1}(a^{t-k-1}b^k)\sum_{\substack{r \in R_{k,t,s}\\ g^{-1}r = \mathbb{e}}}\sum_{\rho \in \hat{G}}d_{\rho}\chi_{\rho}(\mathbb{e}) = \frac{1}{\sqrt{d}N}\sum_{k=0}^{t-1}(a^{t-k-1}b^k)\sum_{\substack{r \in R_{k,t}\\ g^{-1}r = \mathbb{e}}}N\\
     &=  \frac{1}{\sqrt{d}}\sum_{k=0}^{t-1}(a^{t-k-1}b^k)\#_{k,t,s}(g)
\end{align*}
\end{proof}
\subsection{When $S$ is Conjugate Invariant}
Recall that a generating set $S$ is \emph{conjugate invariant} if it
is a union of one or more conjugacy classes.
\begin{corollary}\label{thm: uni conjugate}
If the generating set $S$ is conjugate invariant then  the walk is uniform over the conjugacy classes of $G$. Specifically, the distribution $P_t[\ |_{\psi_0}]$ after $t$ steps  is a class function.
\end{corollary}

\begin{proof}
Suppose the elements $g,h$ are from the same conjugacy class. Let $h = \tau g \tau^{-1}$ for some $\tau \in G$.
Then,
\begin{align}\label{eq: uni  con eq}
    \alpha_{s,t}(h) = \alpha_{s,t}(\tau g \tau^{-1}) = \frac{1}{\sqrt{d}}\sum_{k=0}^{t-1}(a^{t-k-1}b^k)\#_{k,t,s}(\tau g \tau^{-1})
\end{align}
We note that for any $\tau \in G$ the function $\tau^{-1}()\tau : S \to S$ is an automorphism. This implies it is also an isomorphism from $R_{k,t,s}$ to $R_{k,t,\tau^{-1}s\tau}$. 
To show this take $r = s_{k}^{p_k}s_{k-1}^{p_{k-1}}\cdots s_{1}^{p_1}s$. 
Then,
\begin{align}\label{eq: conjugacy aut}
    \tau^{-1}r\tau = \tau^{-1}s_{k}^{p_k}\tau \tau^{-1}s_{k-1}^{p_{k-1}}\tau\cdots \tau^{-1}s\tau = {s'}_{k}^{p_k}{s'}_{k-1}^{p_{k-1}}\ldots {s'}_{1}^{p_1}\tau^{-1}s\tau = r' \in R_{k,t,\tau^{-1}s\tau}
\end{align}
where ${s'}_i = \tau^{-1}s_i\tau$. The last containment follows from the fact that $s_{i} \ne s_{i+1} \iff {s'}_{i} \ne {s'}_{i+1}$. To show injectivity we note that  $\tau^{-1}r\tau = \tau^{-1}r'\tau \implies r = r'$.
Then,
\begin{align*}
    \#_{k,t,s}(\tau g \tau^{-1}) = |\{r \in R_{k,t,s}\mid r = \tau g \tau^{-1}\}| = |\{r \in R_{k,t,\tau^{-1}s\tau}\mid r =  g\}| = \#_{k,t,\tau^{-1}s\tau}(g)
\end{align*}
Substituting the above in Eq. \ref{eq: uni  con eq} we have,
\begin{align*}
    \alpha_{s,t}(h) =  \frac{1}{\sqrt{d}}\sum_{k=0}^{t-1}(a^{t-k-1}b^k)\#_{k,t,\tau^{-1}s\tau}(g) = \alpha_{\tau^{-1}s\tau,t}(g)
\end{align*}
Finally,
\begin{align*}
     P_t[\tau g \tau^{-1}|_{\psi_0}] = \sum_{\tau^{-1}s\tau \in S}{|\alpha_{\tau^{-1}s\tau,t}(g)|}^2 = \sum_{s \in S}{|\alpha_{s,t}(g)|}^2 = P_t[g|_{\psi_0}].
\end{align*}
\end{proof}
\begin{remark}
From the above it follows that the time average distribution $\overline{P}[\mid \psi_0]$ is also a class function.
\end{remark}
\subsection{When $\ket{\psi_0}$ is a Basis State}
In order to determine the mixing time we want to know the distribution starting from a basis state; that is $\ket{\psi_0} = \ket{s_{*},g_{*}}$. $R_{s_*}$ be the set of generating sequences beginning with $s_{*}$.
We define $R_{k,t,s,+s_*} = R_{k,t,s}\cap R_{s_*}$ and $R_{k,t,s,-s_*} = R_{k,t,s}\setminus R_{k,t,s,+s_*}$. Analogous to Eq. \ref{eq: pathcounts} we define,
\begin{align*}
    \#_{k,t,s,+s_*}(g) = |\{\mbox{$r \in R_{k,t,s,+s_*}$}\mid \mbox{$r = g$}\}| \mbox{ and }\#_{k,t,s,-s_*}(g) = |\{\mbox{$r \in R_{k,t,s,-s_*}$}\mid \mbox{$r = g$}\}|
\end{align*}

\begin{theorem}\label{thm: single state}
Starting at $\ket{\psi_0} = \ket{s_*,g_*}$ we have,
\begin{align*}
    \alpha_{s,t}(g) = \sum_{k=0}^{t-1}a^{t-k-1}b^k\left(a\#_{k,t,s,+s_*}(g_*^{-1}g)+b\#_{k,t,s,-s_*}(g_*^{-1}g)\right)
\end{align*}
\end{theorem}

\begin{proof}
The proof is similar to Theorem \ref{thm: main} except the initial step which leads to a dependency on $g_*, s_*$.
\end{proof}
Taken together, the following two lemmas show that, up to a permutation of $G$, the distribution does not depend on the initial state $\ket{s_*,g_*}$, if $S$ is conjugate invariant.
\begin{lemma}\label{lem: perm on g}
\begin{align*}
    P_t[g |_{\ket{s_*,\pi g_*}}] = P_t[\pi^{-1} g |_{\ket{s_*,g_*}}]
\end{align*}
\end{lemma}
\begin{proof}
Since $\#_{k,t,s,\pm s_*}((\pi g_*)^{-1}g) = \#_{k,t,s,\pm s_*}(g_*^{-1}\pi^{-1}g)$.
\end{proof}
\begin{lemma}\label{lem: perm on s}
If the generating set $S$  is conjugate invariant and $s_* \ne s_*'$, then
\begin{align*}
    P_t[g |_{\ket{s_*', g_*}}] = P_t[\pi g |_{\ket{s_*,g_*}}]
\end{align*}
for some  $\pi$ acting on $G$.
\end{lemma}
\begin{proof}
Every generator has the same order and creates cycles of the same length in $\Gamma$ (if $g = \tau h \tau^{-1}$ and $g^k = \mathbb{e}$ then $h^k = \mathbb{e}$). Thus $\Gamma$ is symmetric with respect to its generators. Specifically, the chirality of the initial state $\ket{s_*',g_*}$ specifies the initial ``direction" of the walk. Previous argument implies that these directions are symmetric. Hence, the distribution of the walk is same as when starting from $\ket{s_*,g_*}$ up to a permutation on the vertices of $\Gamma$. More formally, using an  argument similar to that in Corollary \ref{thm: uni conjugate} we can show $\#_{k,t,s,\pm s_*'}(g) = \#_{k,t,s, \pm s_*}(\tau g\tau^{-1})$, where $s_*' = \tau s_* \tau^{-1}$.

\end{proof}
\begin{remark}
Unfortunately, a result analogous to Corollary \ref{thm: uni conjugate} does not hold in this case even if we relax our definition of a class function as follows.
We say, $f$  is a class function up to some permutation iff there exists some fixed permutation $\pi$ acting on $G$ such that: $f(\pi g) = f(h)$ whenever $g$ and $h$ belong to the same conjugacy class. 
The following graph serves as a counterexample.  Let $G = {\cal S}_4$ and $S = $ set of all transpositions of $G$. 
$G$ has $5$ conjugacy classes. However, the probability distribution after the first two steps of the walk starting from $\ket{(1,2),\mathbb{e}}$ has $6$ distinct values:
\begin{align*}
    P_t[ |_{\ket{(1,2),\mathbb{e}}}] = \left(\frac{4}{9}, 0, 0, 0, 0, 0, 0, \frac{2}{27}, \frac{2}{27}, \frac{2}{27}, \frac{1}{27}, \frac{2}{27}, \frac{1}{27}, \frac{1}{27}, \frac{1}{27}, 0, 0, 0, 0, 0, 0, \frac{5}{81}, \frac{2}{81}, \frac{2}{81}\right)
\end{align*}
\end{remark}



\ifx false
\begin{theorem}
The above theorem holds even if $\ket{\psi_0} = \sum_{s,g}\alpha_{0,t}(g)\ket{s,g}$ is an arbitrary starting state. 
\end{theorem}
\begin{proof}
The proof uses a similar analysis as above to represent the amplitudes as a sum of characters.
More specifically we can write,
\begin{align*}
    \alpha_{s,t}(g) = \sum_{g' \in G}\sum_{k=0}^{t-1}f(\psi_0,s,a,b,k)\sum_{r \in R_k}\#_{k,t}({g'}^{-1}gs^{-1})
\end{align*}
where $f(\psi_0,s,a,b,k)$ is a function that does not depend on $g$. An argument essentially identical to  Theorem \ref{thm: uni conjugate} can be used to show that $P_t[\ |_{\psi_0}]$ is a class function.
\end{proof}
\fi

\begin{figure}[h]
	\includegraphics[width=17cm]{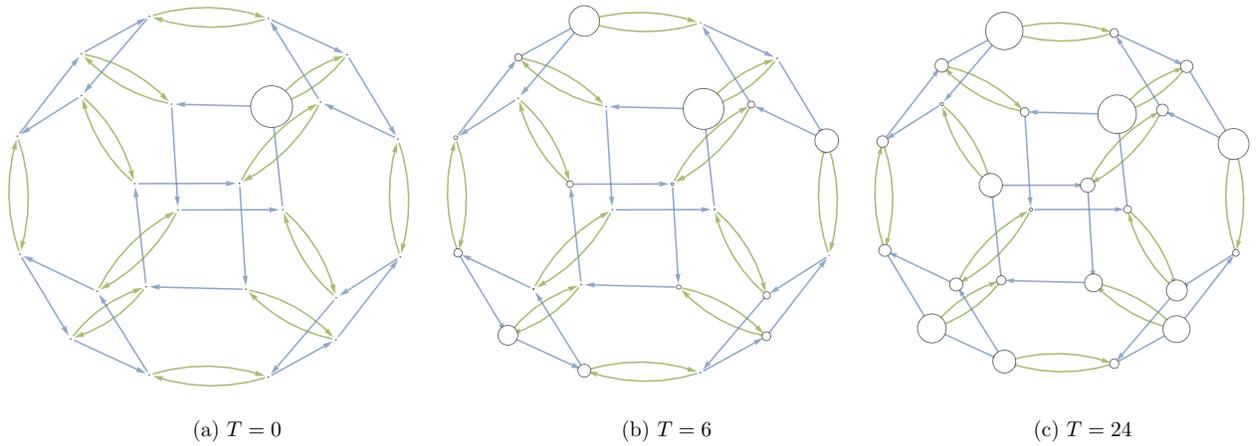}
	\centering
	\caption{Time averaged distribution ($\overline{P}_T[\ \mid_{\ket{0,\mathbb{e}}}]$) of the Hadamard walk on $\Gamma_4$. Vertices are sized proportional to the probability of observing the particle there.}
\label{fig: dist 5}
\end{figure}
\section{The Hadamard Walk on $\Gamma_n$}
In this sections we study the case when the size of the generating set is 2. 
Theorem \ref{thm: main -1} does not apply here directly.
In fact we consider a specific case when $S = \{\mu = (12), \sigma = (1\cdots n)\}$ and $C = H$ is the Hadamard operator. However, the principle techniques used here applies to any arbitrary $C$ and any $S$ with $|S| = 2$. 
In what follows we identify the basis vector corresponding to $\mu$ (resp. $\sigma$) as $\ket{0}$ (resp. $\ket{1}$). 
We can represent a generating sequence $\mu^{p_1}\sigma^{q_1}\ldots\mu^{p_l}\sigma^{q_l}$ , where each $p_i, q_i$'s are non-negative integers, as a $L = \sum_i{(p_i+q_i)}$ bit number $k \in [2^L]$\footnote{$[2^L] = \{0,\ldots,2^L-1\}$}. For example $\mu^2\sigma\mu^3\sigma^2$ is represented as $00100011 = 35$. By $\mu^p$ we represent the sequence $\mu\cdots \mu$, where $\mu$ is applied $p$ times and not the corresponding group element, which is either $\mu$ or $\mathbb{e}$.
Henceforth we identify $\mu$ (resp. $\sigma$) with $0$ (resp. $1$). We use  $\hat{k}$ to  denote the group element corresponding to the generating sequence $k$.
We define a sequence $W_{n}$ of length $2^{n}$ over the alphabet $\{1,-1\}$. Let  $\underline{W}_n$ and $\overline{W}_n$ be the first and last half of $W_n$ respectively. Let $-W_n$ be the negation of $W_n$ ($\forall i -W_n(i) = (-1)W_n(i)$). Then,
\begin{align*}
    W_n = \begin{cases}
    [1,1] \hspace{4.2cm} \mbox{if $n = 1$}\\
    [W_{n-1}\ \underline{W}_{n-1}\ -\overline{W}_{n-1}] \hspace{1cm} \mbox{otherwise}
    \end{cases}
\end{align*}
Loosely speaking, $W_n$'s can be thought of as a vector analogue of the corresponding Walsh matrix.

\begin{theorem}
If $S = \{0, 1\}$ and $C = H$ then starting from the initial state $\ket{\psi_0} = \ket{0,\mathbb{e}}$ the amplitude after $t \ge 1$ steps is given by,
$$\alpha_{s,t}(g) = \frac{1}{\sqrt{2^{t}}}\sum_{\substack{k \in [2^{t}]\\ k = \delta_{s1} \mod 2}}\delta_{\hat{k} , g}{W_{t}(k)}$$
\end{theorem}
\begin{proof}
First we show for $t \ge 1$, 
\begin{align*}
  \ket{\psi_t} = \frac{1}{\sqrt{2^t}}\left(\sum_{\substack{k \in [2^t]\\ k = 0 \mod 2}} W_{t}(k) \ket{0,\hat{k}}+\sum_{\substack{k \in [2^t]\\ k = 1 \mod 2}} W_{t}(k) \ket{1,\hat{k}}\right)
\end{align*}
The proof is via induction. The base case $t = 1$ is trivial.
Applying the Hadamard walk operator to $\ket{\psi_{t}}$ yields,
\begin{align}\label{eq: hadamard}
 \nonumber \ket{\psi_{t+1}} &= \frac{1}{\sqrt{2^{t+1}}}\sum_{\substack{k \in [2^t]\\ k = 0 \mod 2}} (W_{t}(k) \ket{0,\hat{k}0} + W_{t}(k) \ket{1,\hat{k}1})\\ \nonumber &+ \frac{1}{\sqrt{2^{t+1}}}\sum_{\substack{k \in [2^t]\\ k = 1 \mod 2}} (W_{t}(k) \ket{0,\hat{k}0}-W_{t}(k) \ket{1,\hat{k}1})\\
  & = \frac{1}{\sqrt{2^{t+1}}}\left(\sum_{\substack{k \in [2^{t+1}]\\ k = 0 \mod 2}} W_{t+1}(k) \ket{0,\hat{k}}+\sum_{\substack{k \in [2^{t+1}]\\ k = 1 \mod 2}} W_{t+1}(k) \ket{1,\hat{k}}\right)
\end{align}
Where the last equality follows from the definition of $W_t$.
The terms in Equation \ref{eq: hadamard} that contributes towards $\alpha_{s,t}(g)$ are those for which $\hat{k} = g$. This immediately implies the theorem.
\end{proof}

\begin{remark}[Spectra of $U$]
A brief remark about the spectrum  of $U = \Lambda (H \otimes I)$, where $H$ is the Hadamard operator. The case with $C = \frac{1}{\sqrt{2}}(I + iX)$ is similar.
Let $P_\mu$ and $P_\sigma$ be the permutation matrices corresponding to $\mu$ and $\sigma$ respectively. It is an easy exercise to show that $U = \frac{1}{\sqrt{2}}\begin{bmatrix} P_\mu & P_\mu\\  P_\sigma & -P_\sigma
\end{bmatrix}$. Unfortunately, the eigenvalues of $U$ are not all distinct. Hence the minimum eigenvalue gap is zero and we cannot directly use Theorem 6.1 in \cite{aharonov2001quantum} to bound the mixing time.
\end{remark}

\ifx false
Before we compute $f(a,b,t)$ we perform the inverse Fourier transform to get back the amplitudes.
\begin{align}\label{eq: congu}
\nonumber    \alpha_{s,t}(g) &= \frac{1}{\abs{G}}\sum_{\rho \in \hat{G}}d_{\rho}Tr(\rho^{\dagger}(g)\hat{\alpha}_{s,t}(\rho))\\
\nonumber    &=\frac{1}{\abs{G}}\sum_{\rho \in \hat{G}}d_{\rho}Tr(\rho(g^{-1})f(a,b,t)\rho(s^t))\\
    &=\frac{f(a,b,t)}{\abs{G}}\sum_{\rho \in \hat{G}}d_{\rho}Tr(\rho(g^{-1}s^t))
\end{align}

Let $[g]$ be the conjugacy class of $g$. 
Let $h \in [g]$ and $g = khk^{-1}$ for some $k \in G$.
Then
\begin{align*}
    Tr(\rho(g^{-1}s^t)) &= Tr(\rho(kh^{-1}k^{-1}s^{t})) \\ & =Tr(\rho(kh^{-1})\rho(k^{-1})\rho(s^t))=Tr(\rho(kh^{-1})\rho(s^t)\rho(k^{-1}))\\
    &=Tr(\rho(k)\rho(h^{-1}s^t)\rho(k^{-1})) = Tr(\rho(h^{-1}s^t))
\end{align*}
Hence equation \ref{eq: congu} gives the amplitude for the entire conjugacy class $[g]$,  
\fi

\ifx false
Let $g = khk^{-1}$. Then we have,
\begin{align*}
    Tr\left(\rho((khk^{-1})^{-1}g's^t)\right) &= Tr\left(\rho(kh^{-1}k^{-1}g's^t)\right) \\
    &= Tr\left(\rho(kh^{-1})\rho(k^{-1})\rho(g's^t)\right)\\
    &=Tr\left(\rho(kh^{-1})\rho(g's^t)\rho(k^{-1})\right) = Tr\left(\rho(h^{-1}g's^t)\right)
\end{align*}
Hence $\alpha_{s,t}$ is a class function regardless of the initial state of the walk.
This implies for all $s,t$ and $\rho$,

\begin{align*}
    \hat{\alpha}_{s,t}(\rho) =  \lambda_{s,t}(\rho)I
\end{align*}

where $\lambda_{s,t}(\rho) \in \mathbb{C}$ is a constant.
Specifically,

\begin{align*}
    \lambda_{s,t}(\rho) = \frac{\abs{G}}{d_{\rho}}\braket{\alpha_{s,t}}{\chi_\rho^*}
\end{align*}
Substituting this in Eq. \ref{eq: main alpha} we have,

\begin{align*}
    \alpha_{s,t}(g) = \frac{1}{\abs{G}}\sum_{\rho \in \hat{G}}d_{\rho}\sum_{g'}\lambda_{s,t}(\rho)Tr\left(\rho(g^{-1})\right)
\end{align*}

Since $\lambda_{s,t}(\rho)$'s are the only $t$-dependent term in the expression for $\alpha_{s,t}(g)$ it is sufficient to study the convergence of $\lambda_{s,t}(\rho)$'s separately as $t \to \infty$.

Let $P_t[g\mid \psi_0]$ be the probability of observing the particle at $g$ after $t$ steps, starting from the initial state $\ket{\psi_0}$. Then,
\begin{align*}
    P_t[g\mid \psi_0] &= \sum_{s}\abs{\braket{s,g}{\psi_t}}^2 = \sum_s \abs{\alpha_{s,t}(g)}^2\\
    &= \frac{1}{\abs{G^2}}\sum_s\abs{\sum_{\rho \in \hat{G}}d_{\rho}\sum_{g'}\Phi(s,t,g', \phi_0)\chi_\rho(g^{-1}g's^t)}^2 
\end{align*}
In the above expression the $\chi_\rho()$ terms are periodic, whose periodicity depends on the order of $s$ in $G$. Hence the convergence depends only on the terms $\Phi()$'s. In the next section we will study the convergence of $P_t$ for different $\ket{\phi_0}$ and $\cal S$.


\subsection{$\ket{\phi_0} = \frac{1}{\sqrt{\abs{S}}} \sum_{s}\ket{s,g_0}$ and $S, g_0$ is arbitrary}
In this case the initial probability distribution is given by,
\begin{align*}
    P_0[g] &= 
    \begin{cases}
    0\ \mbox{if $g \ne g_0$}\\
    \sum_{s}\abs{\alpha_{s,t}(g_0)}^2
    \end{cases}
\end{align*}
Now,
\begin{align}\label{eq: uniform s}
    \hat{\alpha}_{s,0}(\rho) = \sum_g \alpha_{s,t}(g)\rho(g) = \frac{\rho(g_0)}{\sqrt{\abs{S}}}
\end{align}

Let us compute $a_{t-1}, b_{t-1}$.
Since $a_0 = a = \frac{2}{\abs{S}}-1$ and $b_0 = b = \frac{2}{\abs{S}}$ we have,

\begin{align*}\label{eq: a b}
    a_1 = \left(\frac{2}{\abs{S}}-1\right)^2 + (\abs{S}-1)\left(\frac{2}{\abs{S}}\right)^2 = 1\\
    b_1 = 2\left(\frac{2}{\abs{S}}-1\right)\left(\frac{2}{\abs{S}}\right) + (\abs{S}-2)\left(\frac{2}{\abs{S}}\right)^2 = 0
\end{align*}
From the above we immediately get $a_r = 1$ and $b_r = 0$ for any $r \ge 1$.
Hence,
\begin{align*}
    \Psi(\rho,s, t,\phi_0) &= a_{t-1}\hat{\alpha}_{s,0}(\rho)+b_{t-1}\sum_{s\ne s'}\hat{\alpha}_{s',0}(\rho)
    = \frac{\rho(g_0)}{\sqrt{\abs{S}}}
\end{align*}
Thus,

\begin{align*}. 
\nonumber    \alpha_{s,t}(g) &= \frac{1}{\abs{G}}\sum_{\rho  \in \hat{G}}d_{\rho}Tr(\rho^{\dagger}(g)\hat{\alpha}_{s,t}(\rho))\\
\nonumber    &=\frac{1}{\abs{G}\sqrt{\abs{S}}}\sum_{\rho \in \hat{G}}d_{\rho}Tr(\rho(g^{-1}s^tg_0)) 
\end{align*}

We see that the amplitudes are fully periodic in $t$ with a period $|s|$, the order of $s$ in $G$. From which it follows that the probability $P_t[g]$ is also periodic. Let us now determine the time averaged distribution $\bar{P}_t[g] = \frac{1}{t} \sum_{i=0}^{t-1}P_i[g]$ which is guaranteed to converge.

\begin{align*}
\bar{P}_t[g] = \frac{1}{t} \sum_{i=0}^{t-1}P_i[g] = \frac{1}{t} \sum_{i=0}^{t-1}
\end{align*}

\subsection{$G = S_n$ and $S$ is the set of transpositions}

\section{Arbitrary Coin Operator}
Let $U$ any unitary acting on ${\cal H}_{\cal G,S}$ satisfying the graph adjacency constraints. 
The unitary $U = S(C \otimes I)$ can be decomposed as the coin operator followed by the shift operator.
Let $\{(\lambda_i, \ket{\nu_i})\}$ be the set of eigenvalue-eigenvector pair of $C$.

That is if $U \ket{s, g} = \sum_{g',s'}\alpha_{s',g'}\ket{s',g'}$ then $\alpha_{s',g'} = 0$ if $g' \not \in $

\fi
\bibliographystyle{splncs04}
\bibliography{ref}
\end{document}